# First-principles electronic structure investigation of $HgBa_2Ca_{n-1}Cu_nO_{2n+2+x}$ with the SCAN density functional

Alpin N. Tatan[1,2], Jun Haruyama[2], and Osamu Sugino[1,2]

[1] Department of Physics, Graduate School of Science, The University of Tokyo, Tokyo 113-0033, Japan

[2] Institute for Solid State Physics, The University of Tokyo, Kashiwa, Chiba 277-8581, Japan

(Dated: May 4, 2022)

We perform first-principles calculation to study the electronic structure of $HgBa_2Ca_{n-1}Cu_nO_{2n+2+x}$ copper oxides up to $n = 6$ for the undoped parent compound ($x = 0$) and up to $n = 3$ for the doped compound ($x > 0$) by means of the SCAN meta-GGA density functional. Our calculations predict an antiferromagnetic insulator ground state for the parent compounds with an energy gap that decreases with the number of $CuO_2$ planes. We report structural, electronic and magnetic order evolution with $x$ which agree with experiments. We find an enhanced density of states at Fermi level at $x \approx 0.25$ for the single-layered compound manifesting in a peak of the Sommerfeld parameter, which recently has been discussed as a possible signature of quantum criticality generic to all cuprates.

# I. INTRODUCTION

The discovery of high temperature superconductivity at around 40 K in $\text{La}_{2-x}\text{Ba}_x\text{CuO}_4$ copper oxides (cuprates) in 1986 has opened a new research frontier in condensed matter physics [1]. New superconductors operating above liquid nitrogen temperature (77 K) were soon found in $\text{YBa}_2\text{Cu}_3\text{O}_7$ [2]. Multilayered compounds such as $\text{Bi}_2\text{Sr}_2\text{CaCu}_2\text{O}_{8+x}$ [3, 4] and $\text{HgBa}_2\text{Ca}_{n-1}\text{Cu}_n\text{O}_{2n+2+x}$ [5, 6] then earned significant interest from the scientific community [7, 8] as adding $\text{CuO}_2$ planes could increase the transition temperature $T_\text{c}$ beyond 100 K. The trilayer compound $\text{HgBa}_2\text{Ca}_2\text{Cu}_3\text{O}_{8+x}$ ($T_\text{c}$ = 135 K at ambient pressure [6] and $T_\text{c}$ = 164 K at 31 GPa [9, 10]) is still the highest $T_\text{c}$ superconductor outside the electron-phonon mechanism of Bardeen-Cooper-Schrieffer (BCS) theory [11].

Besides its $T_\text{c}$ value, the Hg-based family (Fig. 1) is preferred for its structural simplicity [12, 13]. Its tetragonal structure is relatively free from structural phase transitions unlike for example, $\text{La}_2\text{CuO}_4$. Its $\text{CuO}_2$ planes have minimal buckling, and there are no Cu-O chains in its cell structure unlike in $\text{YBa}_2\text{Cu}_3\text{O}_7$ [17, 18]. When doped, the dopant O atom is found to reside in the center of Hg layer, i.e., at the $\left(\frac{1}{2}, \frac{1}{2}, 0\right)$ position [19, 20]. From theoretical perspective, Ref. [21] has also affirmed with atomistic simulations that this dopant location is energetically most favorable. The disorder effects to the $\text{CuO}_2$ plane from the dopant are minimized with the thick Ba layer between the Hg and $\text{CuO}_2$ planes (see Fig. 1) [12, 22]. Despite the difficulty in obtaining large single crystals with well-defined composition [23], these materials serve as ideal benchmarks for assessing theoretical models.

Unlike other prototypical cuprates such as $\text{La}_2\text{CuO}_4$, there is no experimental data for the electronic band gap and other electronic structure features in the parent compound $\text{HgBa}_2\text{Ca}_{n-1}\text{Cu}_n\text{O}_{2n+2}$ due to the difficulty to synthesize high-purity samples. Thus, its description is both a challenge and opportunity for first-principles calculation techniques. Early studies based on local-density or generalized gradient approximations (LDA/GGA) to density functional theory (DFT) were insufficient because an incorrect nonmagnetic metallic ground state was predicted instead of the antiferromagnetic Mott insulating ground state [24 – 27]. The more sophisticated hybrid exchange-correlation functionals that include a fraction of the nonlocal Fock exchange could provide a reasonable description of the ground state. For example, the HSE06 functional [28] opens a finite insulating gap and predicts a magnetic exchange coupling ($J$) value that is comparable to experiment in the antiferromagnetic ground state of $\text{La}_2\text{CuO}_4$ [29]. However, the hybrid methodology fails to describe the insulator-to-metal transition upon doping for lanthanum cuprates [30, 31] as the energy gap persists even in the doped state. A suitable exchange-correlation functional that can properly describe both the undoped and doped phases of cuprates remain an important research focus until this day.

The multilayered parent compounds ($\text{HgBa}_2\text{Ca}_{n-1}\text{Cu}_n\text{O}_{2n+2}$) have been studied using hybrid functionals for $n = 1, 2, 3$ (hereafter, we refer to these as Hg-12$(n-1)n$, e.g., Hg-1201, Hg-1212, and Hg-1223 for $n = 1, 2, 3$ respectively) [27]. The obtained magnetic solutions concurred with other cuprates, e.g., $\text{La}_2\text{CuO}_4$. The magnetic exchange coupling was slightly overestimated compared to configuration interaction calculations with cluster model (CICM) [32, 33]. The density of states (DOS) calculations concluded that unlike the monolayer Hg-1201 which is an insulator with a band gap slightly above 1 eV, the multilayered Hg-1212 and Hg-1223 compounds are metallic as the Hg-O conduction states lower in the energy and the Cu-O conduction states remain fixed, thereby closing the band gap. As this metallicity is not due to Cu states, this would explain the persisting antiferromagnetism in metallic Hg-1212/Hg-1223.

While the presence of low-lying Hg-O states in the conduction band has been suggested since the early calculations of $HgBa_2Ca_{n-1}Cu_nO_{2n+2}$ [34, 35], the extent of mercury roles in the multilayered compounds remain unclear. There has neither been experimental evidence of the metallicity due to low-lying Hg-O states nor further studies with hybrid functionals reported on these materials. This stagnancy may be attributed to the difficulties in synthesizing pure samples and the prohibitive cost of hybrid methodology (approx. hundreds of times more expensive [36, 37] than using semilocal functionals). Thanks to the wider availability of high-performance computing resources, more extensive calculations have become possible in the past decade. Hence, we would like to investigate this material more comprehensively with the computational resources available at our disposal (see Acknowledgments).

Over the years, there were questions on whether band gaps predicted by density-functional calculations should be compared with experimentally derived energy gaps. These doubts are expected to fade since modern computational packages such as VASP [38, 39] have implemented their calculations in the generalized Kohn-Sham (gKS) scheme. Ref. [40] has shown that the gKS band gap is equal to the fundamental band gap in the solid, which is defined as the ground-state energy difference between systems with different numbers of electrons. This provides a solid basis for comparing band gaps of gKS formalism with the experimentally observed band gaps and improvement in the prediction of energies and structures brought by a functional would also indicate improved gKS band gaps [31, 41–43].

The band gap prediction of $La_2CuO_4$ with hybrid methodology [29] served as one of the supporting basis for calculating $HgBa_2Ca_{n-1}Cu_nO_{2n+2}$ in Ref. [27]. Citing the 2 eV optical absorption peak in Ref. [44], the computed $La_2CuO_4$ band gap of 2.5 eV by HSE06 hybrid functional was considered to concur with experiment. However, Ref. [45] also reported a smaller band gap of 0.89 eV obtained via Hall transport measurements. It has since been argued [31, 42, 45, 46] that one should estimate the band gap not from the lowest energy absorption peak, but from the leading-edge gap in the optical spectra. This implies that the computed band gaps by density functional theory should be compared to an experimental value around 1 eV from Ref. [44]. From a comparative study [31], it is apparent that the hybrid functionals overestimate the band gap of $La_2CuO_4$ and another functional of the meta-GGA class is better suited for this purpose, which we shall explain in the following paragraph.

The recently devised strongly-constrained-and-appropriately-normed (SCAN) meta-GGA exchange-correlation functional [37] satisfies all 17 known constraints applicable to meta-GGA. It has successfully described the properties of pristine and doped $La_2CuO_4$ [42, 43], $YBa_2Cu_3O_6$ [46], and $Bi_2Sr_2CaCu_2O_{8+x}$ [47]. In $La_2CuO_4$, SCAN obtains the magnetic moment in magnitude and orientation, the magnetic exchange coupling strength $J$, the magnetic form factor as well as the electronic band gap that well-correspond to experiments. Ref. [31] compared SCAN with 12 other functionals spanning across the levels of Perdew-Schmidt hierarchy [48] in lanthanum cuprates and demonstrated SCAN's superiority in matching experimental results. Although hybrid functionals' value of magnetic exchange coupling $J = 187$ meV [29] is in reasonable agreement with experimental value of $J = 133 \pm 3$ meV [49], a much closer prediction $J = 138$ meV can be obtained with SCAN [42]. In doped $YBa_2Cu_3O_7$, the charge, spin and lattice degrees of freedom are treated equally in a self-consistent manner [46] to yield stable stripe phases without invoking free parameters, which leads to the identification of a landscape of 26 competing uniform and stripe phases. These results indicate that SCAN captures many key features of the cuprates and provides a new prospect in describing correlated electron properties of these materials. The computational cost of meta-GGA functionals that is only a few times larger than its LDA/GGA predecessors adds to the viability of more effective studies of cuprates in comparison to the cost-prohibitive hybrid methodology or beyond-DFT techniques.

In this work, we present a SCAN density-functional description for the electronic structure of $HgBa_2Ca_{n-1}Cu_nO_{2n+2}$ series. Using relaxed structures that are in good agreement with experiment, we show that these compounds remain insulating up to $n = 6$ with a finite but decreasing, indirect band gap. The low-lying Hg-O conduction states are noted, but their dominant proportions against the Cu-O states are not apparent in Hg-1212/Hg-1223 and these Hg-O states are only clearly lower in energy at $n = 6$. In addition, we also investigate the doped phase via supercell construction of Hg-1201, Hg-1212, and Hg-1223 for several representative excess oxygen levels $x$. We report that SCAN improves an earlier description of the doped phase with semilocal functionals [50] in capturing lattice contraction and the magnetic moment as a function of $x$. We confirm the expected narrow DOS peak due to additional states contributed by the dopant O atom at low doping levels and extract an optimum doping $x_o$ where this feature is located at the Fermi level $E_F$. Finally, we compute the normal-state, zero-temperature Sommerfeld parameter of electronic specific heat $\gamma$ from the DOS at $E_F$ and observe a peak across doping levels to support an experimental prediction of it being a universal property among cuprates which could be a signature of quantum criticality [51–53].

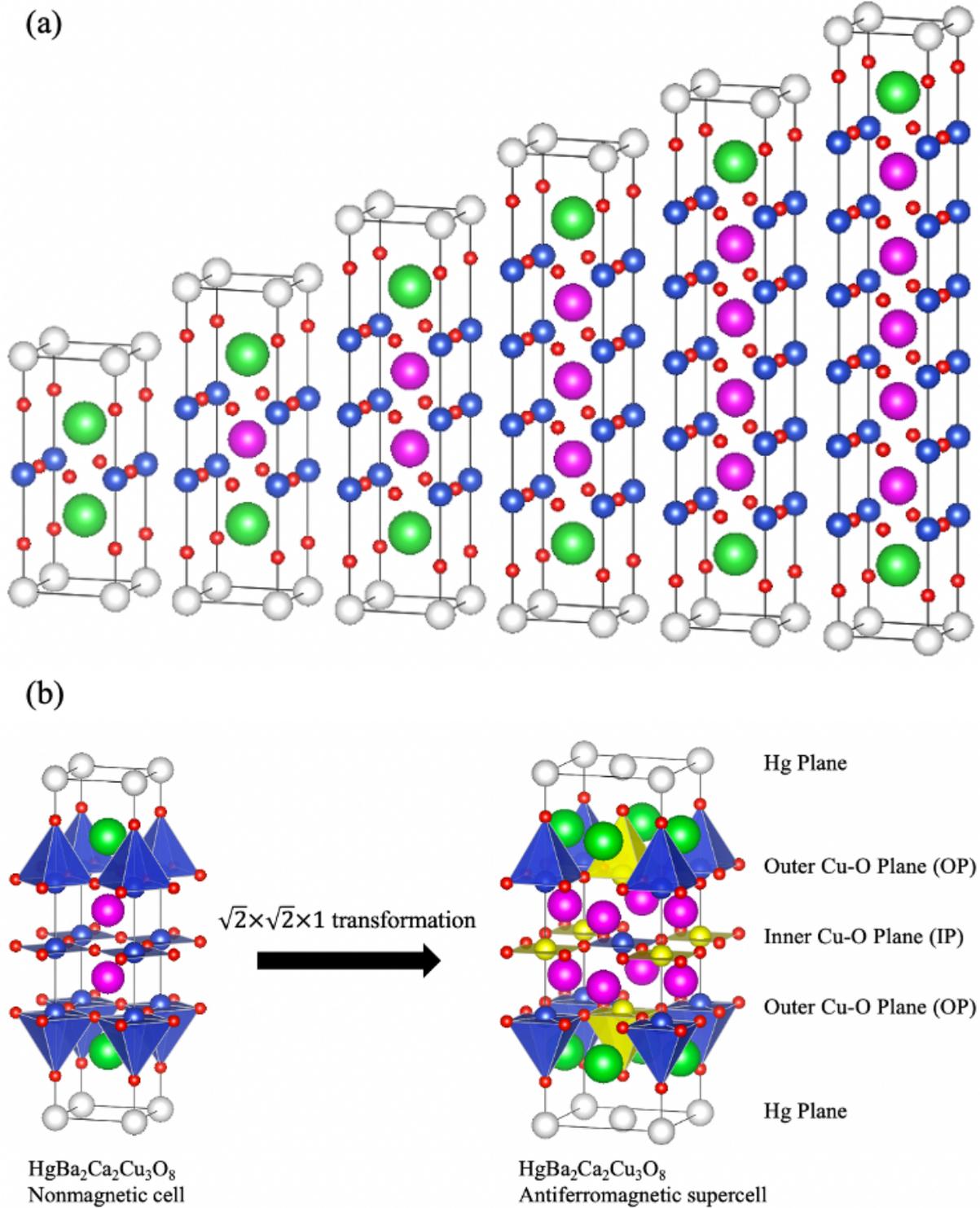

**Figure 1**. (a): A schematic of $HgBa_2Ca_{n-1}Cu_nO_{2n+2}$ structure up to $n = 6$. The white, green, magenta, blue and red colored spheres represent Hg, Ba, Ca, Cu, and O atoms, respectively. (b): The trilayer $HgBa_2Ca_2Cu_3O_8$ compound in its nonmagnetic conventional unit cell and antiferromagnetic $\sqrt{2} \times \sqrt{2} \times 1$ supercell. The copper-oxide and mercury planes are shown. The different colors of Cu-O polyhedra represent opposing magnetic moments in the antiferromagnetic case. Figure prepared with VESTA [14] with atomic positions derived from *Materials Project* database [15] and Ref. [16].

## II. COMPUTATIONAL DETAILS

In general, we followed the computational parameters of the preceding SCAN studies of other cuprates [42, 43, 46, 47]. Ab initio calculations were carried out by using the pseudopotential projector augmented-wave (PAW) method [54] implemented in the Vienna *ab initio* simulation package (VASP [38, 39]) with an energy cutoff of 550 eV for the plane-wave basis set. Exchange-correlation effects were treated using the SCAN meta-GGA scheme [37]. The crystal structures were relaxed using a quasi-Newton (RMM-DIIS) algorithm to minimize energy with an atomic force tolerance of 0.008 eV/Å and a total energy tolerance of $10^{-5}$ eV. The costlier conjugate gradient algorithm was also utilized in a few cases when the forementioned algorithm encountered convergence problems. The relaxation procedure utilized an $8 \times 8 \times 4$ gamma-centered k-point mesh to sample the Brillouin zone. Denser meshes ($20 \times 20 \times 4$ or more) were used to calculate the DOS with tetrahedron method with Blöchl corrections. The band structures are drawn along the path $\Gamma - X - M - \Gamma - Z - R - A - Z$ in the antiferromagnetic $(\sqrt{2} \times \sqrt{2} \times 1)$ cell of the first Brillouin zone. The DOS and bandstructure plots are made with PyProCar [55] and Sumo [56] packages.

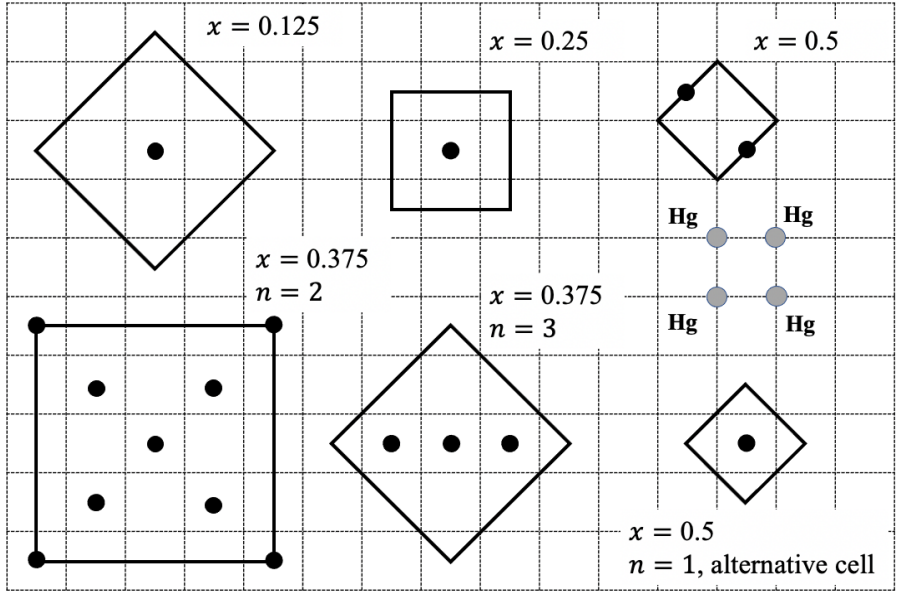

**Figure 2**: Basal planes of doped supercells computed in this study viewed from the top. The dopant oxygen atoms (black spheres) are placed between the mercury atoms (grey spheres). A unit dotted square represent a unit cell meanwhile solid lines represent supercells for each excess oxygen concentration $x$. The top row illustrates the supercells used in all three compounds. The bottom row shows the additional supercells used to compute $x = 0.375$ in Hg-1212 and Hg-1223, as well as an alternative supercell for $x = 0.5$ tested for Hg-1201.

For investigating the doping-dependent electronic structure, we used a series of supercells containing excess oxygen atoms. We used the same calculation parameters for computing both undoped and doped compounds, except for the additional dopant atoms. Oxygen concentrations of $0.125, 0.25, 0.5$ with cell sizes up to an eightfold single cell are considered for compounds Hg-1201, Hg-1212, and Hg-1223. The corresponding cells used for all compounds are illustrated in the top row of Fig. 2. The bottom row of Fig. 2 includes supercells used to compute the $x = 0.375$ case for Hg-1212 and Hg-1223. An alternative dopant placement is also tested for $x = 0.5$ case in Hg-1201 (the bottom right figure in Fig. 2), for which we note only minor difference in results that do not change the conclusion of this study. For the larger supercells of the multilayered compounds containing 8 formula units and more, the DOS plots used a smaller 10 × 10 × 4 k-point mesh due to the large memory requirements. This change does not affect the resulting DOS plots given the smaller size of first Brillouin zone for larger supercells. The doping effect on the lattice parameters, atomic positions and magnetic order was investigated by total-energy and atomic-force calculations. Initial antiferromagnetic order was assumed in our structure, which allowed us to study the interplay between doping levels and the strength of magnetic moments. In this regard, our calculation is an extension of Refs. [50, 57] in which their calculations were performed on non-magnetic, non-relaxed supercells of Hg-1201 with local-density approximation.

Using the results of our DOS calculations, we compute a thermodynamic quantity that can be compared with experiments. The Sommerfeld parameter $\gamma$ of electronic specific heat is defined as:

$$C_{el}(T \to 0) = \gamma T \tag{1}$$

$$\gamma = \frac{2}{3}\pi^2 k_B^2 N(E_F) \tag{2}$$

where the DOS at Fermi energy $N(E_F)$ is defined per atom and per one spin direction [17]. This quantity in the normal state extrapolated to zero temperature can be extracted from $N(E_F)$ of a computational cell containing $X$ formula units [17]:

$$\gamma_n^0 \text{ (mJ/K}^2 \cdot \text{mol)} \approx 2.36 \cdot 2N^0(E_F) \text{ (state / eV} \cdot X) \tag{3}$$

which allows us to estimate $\gamma$ directly from the DOS.

## III. RESULTS AND DISCUSSION

### A. Parent compounds

We plot the in-plane lattice parameter $a$ of $HgBa_2Ca_{n-1}Cu_nO_{2n+2}$ structures relaxed with SCAN in Fig. 3. This parameter decreases with the number of copper-oxide planes $n$ in agreement with LDA results [58] and the trend observed in experiments for doped structures. SCAN predicts values that are closer to experiments than LDA. Moreover, SCAN provides slightly larger parameters for the parent compounds compared to experimental values for doped samples. The LDA results, however, are on the smaller side, reflecting its well-known overbinding issue. Our SCAN results are hence an improvement as it agrees with doping-induced lattice contraction observed in experiments. A similar observation can be made for the out-of-plane lattice parameter $c$ (see Fig. S1 and Table S1 of the *Supplemental Material* for the full set of $a$ and $c$ values).

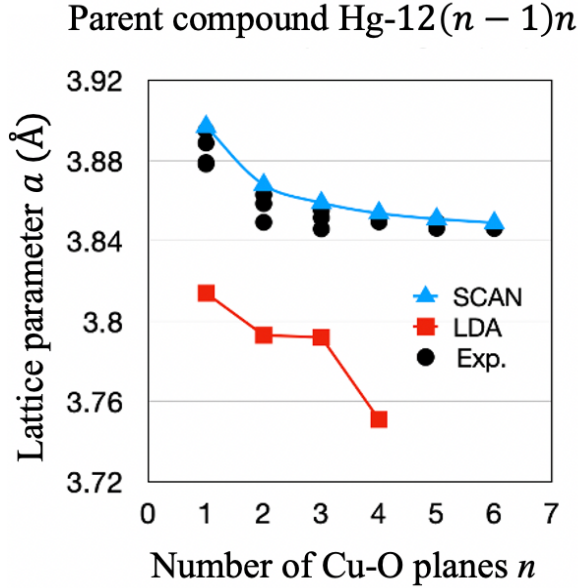

**Figure 3**: In-plane lattice parameter $a$ for Hg-12($n$ − 1)$n$ parent compounds. Triangles and squares denote relaxed parameters obtained with SCAN (this work) and LDA (Ref. [58]). Straight lines are guides to the eye. Black circles are experimental values for doped structures retrieved from Refs. [9, 16, 59–64].

**Table I**: Magnetic moments of Hg-12($n$ − 1)$n$ parent compounds in the antiferromagnetic phase. For $n > 2$, the magnetic moments for the inner planes are shown in parentheses.

| Hg-1201 | Hg-1212 | Hg-1223 | Hg-1234 | Hg-1245 | Hg-1256 |
|---------|---------|---------|---------|---------|---------|
| 0.491   | 0.477   | 0.475 (0.470) | 0.473 (0.469) | 0.472 (0.467) | 0.472 (0.468) |

We tabulate the magnetic moments in the antiferromagnetic ground state of $HgBa_2Ca_{n-1}Cu_nO_{2n+2}$ in Table I. The multilayered compounds have similar magnetic moments of about $0.47\ \mu_B$, with a slight difference between the outer (OP) and inner (IP) planes for $n \geq 3$. These values are comparable with the values predicted for other cuprates by SCAN [42, 43, 47]. The single-layered Hg-1201 compound's magnetic moment agrees with the 0.4 $\mu_B$ value predicted from variational Monte Carlo [65], and the slightly smaller moments for the IP concur with the hybrid functional picture in Ref. [27].

Moving to the electronic structure results, the band gap of $HgBa_2Ca_{n-1}Cu_nO_{2n+2}$ are shown in Fig. 4. In contrast to the metallic multilayered structure predicted by hybrid functionals in Ref. [27], SCAN predicts a decreasing but finite band gap from $n = 1$ to $n = 6$. In addition, our prediction for the monolayer Hg-1201 band gap $E_g^{SCAN} \approx 0.6$ eV is comparable to the results of variational Monte Carlo $E_g^{VMC} \approx 0.7$ eV [65] while the HSE06 and B3LYP hybrid functionals in Ref. [27] yielded much larger band gaps of 1.1 and 1.5 eV, respectively.

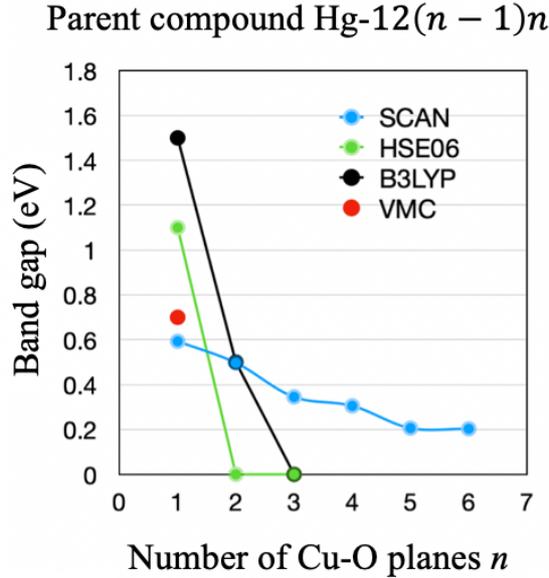

**Figure 4**: Band gap of Hg-12($n − 1$)$n$ parent compounds predicted by SCAN (blue circles, this work), HSE06 and B3LYP hybrids (green and black circles, Ref. [27] ), and variational Monte Carlo (VMC, red circle, Ref. [65]). Straight lines are guides to the eye.

The band structure and DOS plots are shown in Fig. 5 and Fig. 6. These suggest an insulating ground state with indirect band gaps between the *X* and *M* points. The valence bands are dominated by Cu and O contributions as generally expected from cuprates [17]. Starting from $n = 3$, the equivalence between copper oxide planes is broken (Fig. 6 (a)). There is magnetic inhomogeneity between the OP and IP, with the latter having higher energy states. The low-lying, mercury conduction states are not apparently dominant until $n = 6$ in the SCAN description (Fig. 6 (b)). Even at $n = 6$, the structure remains insulating with a finite band gap of $E_g \approx 0.2$ eV. In the SCAN picture, the diminishing band gap is a gradual process, where Cu, Hg, and O conduction states collectively getting closer to the valence states. This is in contrast with the hybrid picture (Ref. [27]) where the insulator-to-metal transition is immediate and is solely facilitated by the Hg states. The full set of DOS plots for the parent compounds are provided in Fig. S2 of the *Supplemental Material*.

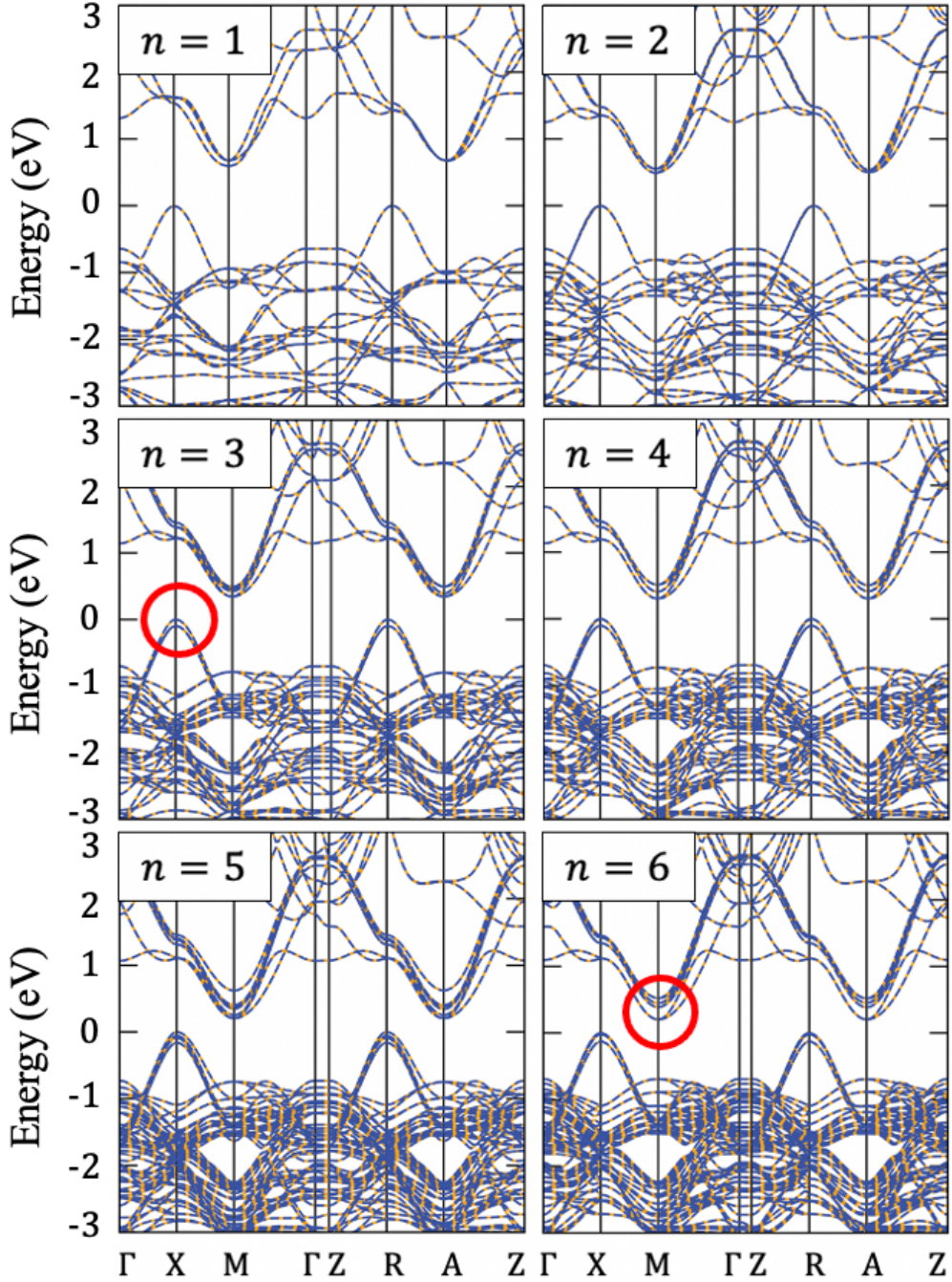

**Figure 5:** Band structure of Hg-12($n-1$)$n$ parent compounds predicted by SCAN for $n=1$ to $n=6$, drawn along the path $\Gamma-X-M-\Gamma-Z-R-A-Z$ in the first Brillouin zone of the magnetic $\sqrt{2}\times\sqrt{2}\times 1$ cell. Red circles are guides to the eye for Cu valence band splitting starting from $n=3$ and the low-lying Hg-O conduction bands at $n=6$. The two spin orientations (blue and yellow traces) coincide and appear as bicolored dashed lines.

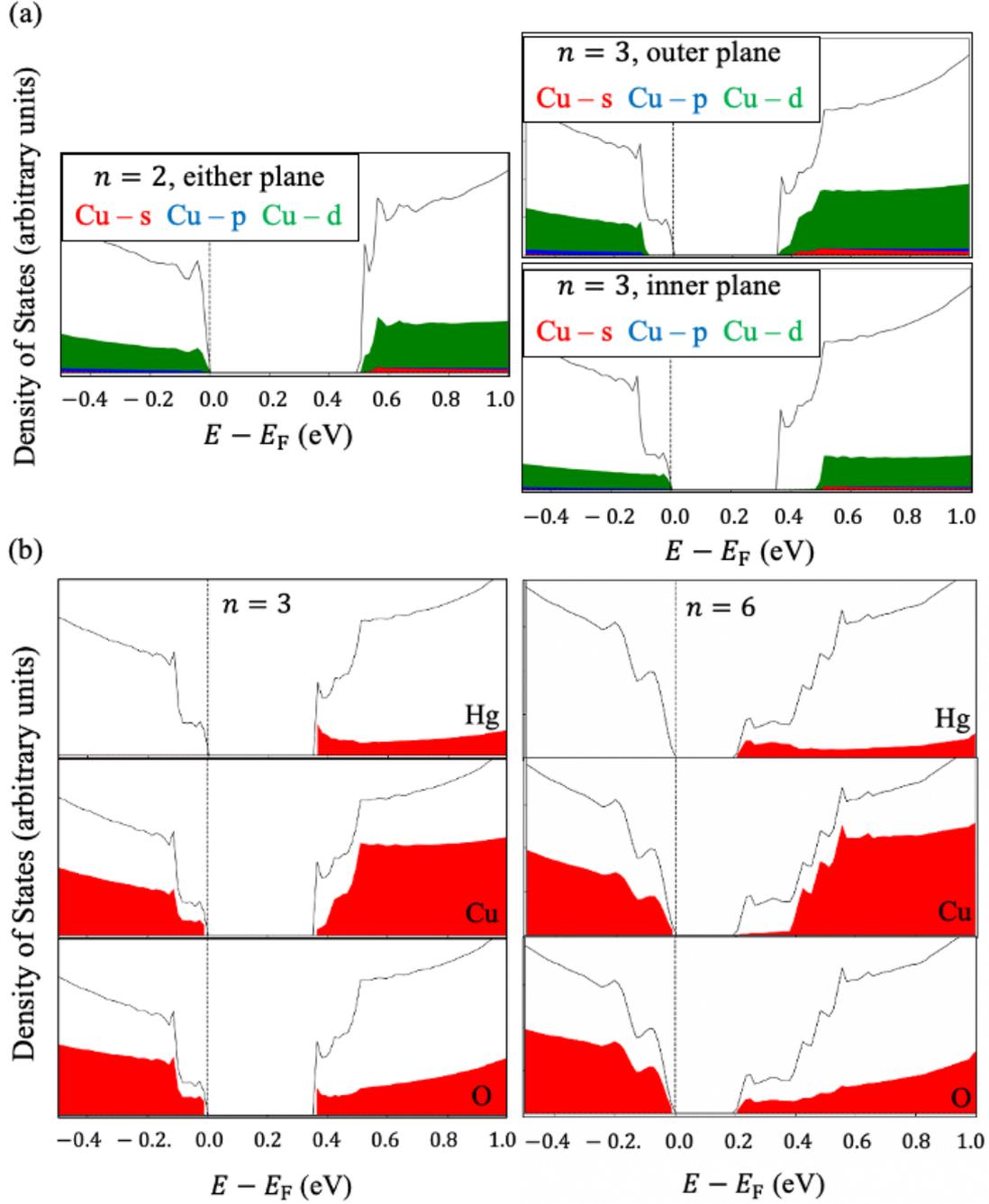

**Figure 6:** DOS plots near $E_F$ for selected Hg-12$(n-1)n$ compounds. Total DOS (black lines, no shading) are presented along with the Cu, Hg, and O contributions (shaded areas). **(a):** the Cu states in equivalent planes of Hg-1212 are shown in contrast to the outer and inner planes of Hg-1223. The colors represent contributions from different Cu orbitals. **(b)**: Hg, Cu and O states in Hg-1223 and Hg-1256.

## B. Doped compounds

We show the lattice parameters of Hg-1201, Hg-1212, and Hg-1223 relaxed with SCAN alongside their experimental values in Fig. 7. Due to limited number of experiments, we collect values over a short range of excess oxygen levels $x$ for each compound from multiple sources. The LDA results [50] for Hg-1201 are included for comparison. Doping-induced lattice contraction is observed in all cases. In the low doping levels ($x \leq 0.125$), there is a good agreement between SCAN (closed circles) and experiments (open circles) for Hg-1201. In comparison to LDA [50], the discrepancy between theory and experiments is improved with SCAN. On the other hand, the experimental lattice contraction is smaller than our results for the multilayered compounds. Ref. [67] noted that their synthesized lattice parameters are very close to the intrinsic size of the $CuO_2$ plane of 3.855 Å in the infinite-layered compound $CaCuO_2$ [68] such that it is difficult to further reduce the lattice size during their synthesis, which may explain the small contraction. Our supercell calculations are not subject to these experimental complexities and thus by varying solely the oxygen concentration we find that this translates to a bigger lattice contraction. We also note that these quantitative discrepancies in the doped lattice parameters are not unexpected because we perform zero-temperature, normal state density-functional calculations while Ref. [67] measured their samples at finite temperatures in the superconducting phase. Nevertheless, we believe capturing qualitatively the doping-induced lattice contraction observed in experiments is still a good progress for theoretical simulations.

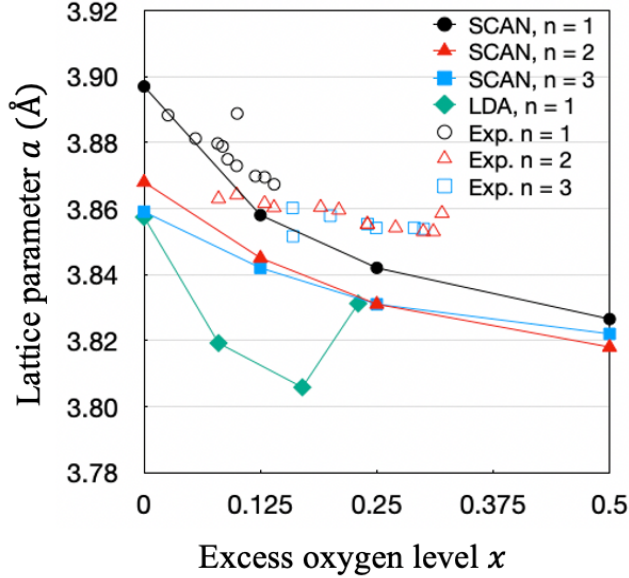

**Figure 7**: In-plane lattice parameter $a$ for Hg-12($n-1$)$n$ doped compounds. Closed (open) circles, triangles, and squares denote the SCAN-computed (experimental) values for $n = 1, 2, 3$ respectively. The experimental values (Refs. [9, 59, 66, 67]) and the LDA-computed values for $n = 1$ from Ref. [50] are included for comparison.

In our doping model that assumes initial antiferromagnetic order, the ground-state magnetic moments are reduced with $x$, as shown in Fig. 8. The single-layered Hg-1201 has the largest drop among all three compounds for the same $x$, which suggests that the reduced magnetic order is related to the hole doping concentration $p$ in each copper-oxide plane. This observation is further affirmed by the results for Hg-1223, where the IP retains more magnetic moments in comparison to the OPs which are closer to the dopant O atom located at the Hg-plane. This magnetic moment inhomogeneity qualitatively agrees with the experiment observation concerning doped five-layered compound Hg-1245 [69], where the IPs retain its antiferromagnetic order ($\mu \approx 0.6\ \mu_B$) in contrast to the much weaker moments in the OPs ($\mu \approx 0.1\ \mu_B$). This demonstrates the capability of SCAN in predicting the charge-spin interplay, which has also been demonstrated in its predictions of doped $YBa_2Cu_3O_7$ [46], and $Bi_2Sr_2CaCu_2O_{8+x}$ [47].

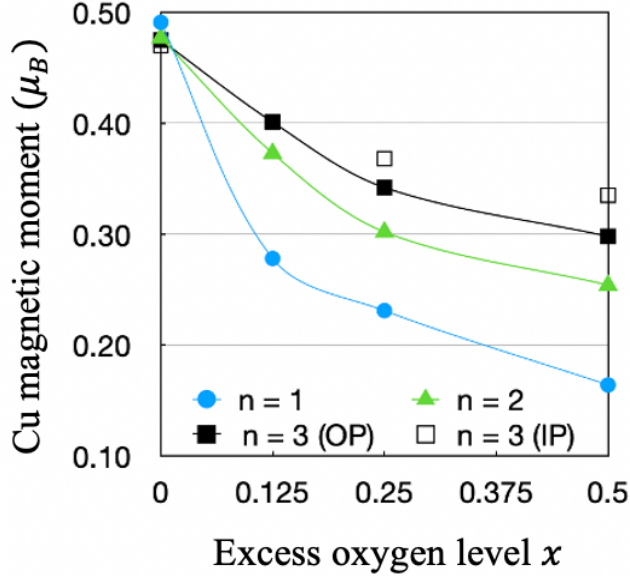

**Figure 8**: Magnetic moments of Hg-12($n − 1$)$n$ doped compounds. The values for single-layered Hg-1201 and bilayered Hg-1212 compounds are denoted by circles and triangles. The outer and inner planes of the trilayered Hg-1223 compound are shown as closed and open squares, respectively.

We analyze the DOS of the doped structures in Fig. 9, where we illustrate the total DOS (black lines) and the contribution from the dopant O atom (shaded area). The latter is localized in a narrow energy range at low concentrations, resulting in a sharp peak of the total DOS at some energy $E < E_F$ (Fig. 9, top). As $x$ is increased, this peak becomes delocalized and its magnitude relative to the other atoms' contributions changes. Considering this evolution of the dopant state, there are two possibilities of its influence on the total DOS at Fermi level $N(E_F)$ for a chosen $x$: first, if the dopant state has a narrow energy range with a magnitude that is larger than other atoms' contributions as it reaches $E_F$, then $N(E_F)$ becomes significantly enhanced (Fig. 9, middle). On the other hand, there will be no induced peak of $N(E_F)$ when the dopant state is highly delocalized with a low magnitude (Fig. 9, bottom). We include similar plots for Hg-1212, and Hg-1223 in Fig. S3 of *Supplemental Material*.

We briefly discuss the nature of the DOS by segregating atomic contributions in Hg-1201 for $x = 0.125$ and $x = 0.25$ in Fig. S4 and Fig. S5 of *Supplemental Material*, respectively. We note a strong similarity from the contributions of dopant O atom with the atoms in its vicinity (Hg, Ba, and apical O atoms) in both Fig. S4 and Fig. S5. At $x = 0.25$, there are also small peaks at $E_F$ for contributions from Cu and planar O, which suggest that the dopant O state delocalizes and interacts with further atoms as the doping level increases. In addition, we remark that these apparent interactions between the dopant O state with other atomic states reported here have also been further elaborated in previous literatures [50, 70–72].

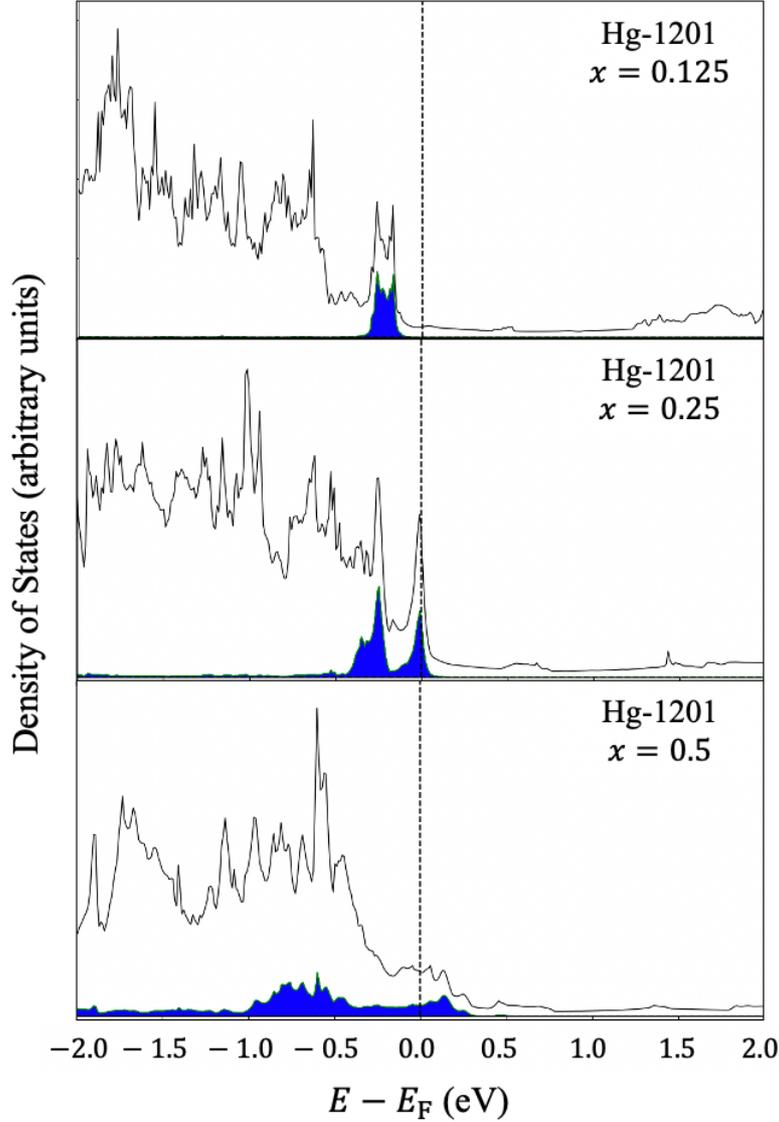

**Figure 9**: Density of states plots for doped $HgBa_2CuO_{4+x}$ in arbitrary units. The total DOS and the contribution from dopant O states are shown in black trace and blue shaded area, respectively. The dopant contribution is sharp and narrow at low excess oxygen level $x$. Its delocalization in energy may result in a peak at the Fermi level at a suitable $x = x_o$ which is approximately equal to 0.25 as shown in the middle plot. At high $x$, the dopant state fully delocalizes and does not lead to a sharp peak in the total DOS.

Our observation in the previous paragraph allows us to define an optimum excess oxygen concentration $x_o$ with regards to the total DOS at $E_F$. For Hg-1201, we can specify $x_o \approx 0.25$ based on the DOS peak observed in Fig. 9. Our few supercell calculations are insufficient to pinpoint the corresponding value for multilayered compounds. However, we can estimate based on the DOS of the sampled concentrations that $0.25 < x_o < 0.375$ for Hg-1212 and $0.375 < x_o < 0.5$ for Hg-1223, forming an increasing trend with the number of copper-oxide layers. This suggests that $x_o$ could be directly related to the amount of hole doping $p$ effectively introduced to each layer. Indeed, this was also the conclusion arrived in Ref. [50] where their optimum value $x^{\text{LDA}} = 0.22$ is attributed to the point where the number of induced holes on the CuO$_2$ plane saturates. The subsequent question is pertaining how we should interpret the physical role of $x_\text{o}$. For example, it is interesting to relate these values with the optimum concentrations that yield the highest superconducting transition temperature, recorded in Refs. [73–75] for these three compounds (Table II). The $x_\text{o}$ for Hg-1201 from DFT calculations obtained in our work and Ref. [50] agree with these records within the experimental uncertainties. Meanwhile, we note that the estimation of optimum doping for $T_\text{c}$ in cuprates is not unequivocal among experiments. In contrast to Refs. [73–75], Refs. [66, 67] predicted a different set of values of $x$ that yield maximum $T_\text{c}$ for these three compounds. This discrepancy may be caused by different experimental techniques used (iodometric titration vs thermogravimetric analysis), which we as theorists claim no expertise of, and yet we may note that these conflicting reports exemplify the long-standing question [35, 70] on whether the induced hole concentration in the copper-oxide plane deviates from a simple ionic picture whereby two holes are donated for every oxygen dopant $x$. The results from Refs. [66, 67] support the simple ionic picture that gives hole concentration $p \approx 2x$, while Refs. [73–75] suggests a smaller dependence $p \approx 0.72x$. Although there is no experimental consensus yet [50], earlier density-functional calculations in Refs. [50, 70] concur with Refs. [73–75]. This is also what we infer from our SCAN calculations. In any case, both ionic and non-ionic pictures suggest that the optimum excess oxygen concentrations for the whole compound follow the ascending order of Hg-1201, Hg-1212, and Hg-1223, which agrees with our results.

Table II: The optimum compounds predicted by SCAN and experiments based on ionic and non-ionic pictures. For SCAN, we define the "optimum" compound to be the one with a sharp DOS peak at $E_\text{F}$ due to contribution of the dopant O states. The experiments' optimum values refer to the compounds that yield maximum superconducting transition temperature.

| "Optimum" Compounds | LDA (Refs. [50, 70]) | SCAN (This work) | Exp. (Ref [67]) (Ionic picture) | Exp. (Ref. [73]) (non-ionic picture) |
|---|---|---|---|---|
| HgBa$_2$CuO$_y$ | $y \approx 4.22$ | $y \approx 4.25$ | $y \approx 4.09$ | $y = 4.18 \pm 0.1$ |
| HgBa$_2$CaCu$_2$O$_y$ | No data | $6.25 < y < 6.375$ | $y \approx 6.21$ | $y = 6.34 \pm 0.12$ |
| HgBa$_2$Ca$_2$Cu$_3$O$_y$ | $y \approx 8.5$ | $8.375 < y < 8.5$ | $y \approx 8.29$ | $y = 8.45 \pm 0.16$ |

We compute the normal-state, zero-temperature Sommerfeld parameter of the electronic specific heat $\gamma$ from the DOS at the Fermi level $N(E_\text{F})$ for the doped compounds in Fig. 10. Our values for the small and large doping levels agree well with the experimental results from other cuprates [17, 51, 52], which lie around $2 - 7$ mJ/K$^2 \cdot$ mol. For Hg-1201, we observe a peak feature across the doping concentration at $x_\text{p} = 0.25$. This feature is of recent interest [51] as it may be a thermodynamic signature of a quantum critical point, indicated by a logarithmic divergence in $C/T \propto \log(1/|x - x^*|)$ where $x$ is some tuning parameter such as the doping concentration. Similar peaks have not been confirmed in our calculations for Hg-1212 and Hg-1223 in Fig. 10 at the concentrations computed in our supercells. Should this feature extend to the multilayered compounds, we expect them to materialize at concentrations $0.25 < x_\text{p} < 0.375$ for the bilayer Hg-1212 and $0.375 < x_\text{p} < 0.5$ for the trilayer Hg-1223 compounds based on the dopant state energies computed in our supercells. The peaks in $\gamma$ have only been experimentally confirmed by direct measurements on lanthanum-cuprate families [51, 76], but there are recent observations on the bismuth- and mercury- cuprate samples that suggest that this is a universal property for all cuprates [52, 53]. However, the exact nature of this divergence is still under debate. Beside the quantum criticality argument, there is an alternative explanation without invoking broken symmetries from two-dimensional Hubbard model [77] which associates the feature to arise from the finite-temperature critical end point of a first-order transition between a pseudogap phase with dominant singlet correlations and a metal. Our result for Hg-1201 is therefore a positive indicator for density-functional methods to contribute, in the future, a deeper theoretical study to understand the nature of this phenomenon.

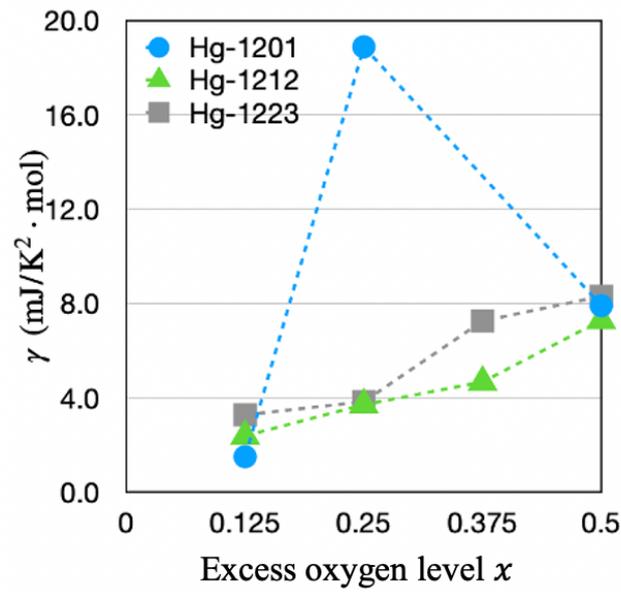

**Figure 10**: The computed zero-temperature, normal-state Sommerfeld parameter $\gamma$ for the Hg-$12(n-1)n$ doped compounds. There's a peak confirmed for the single-layered Hg-1201, while no peak was observed for the multilayered compounds at the excess oxygen levels computed in this study.

## IV. SUMMARY AND CONCLUSIONS

We have studied both the parent and doped compounds of $HgBa_2Ca_{n-1}Cu_nO_{2n+2+x}$ with first-principles calculations utilizing the recently devised SCAN meta- GGA density functional. Our results suggest an improvement in structural characterization of these compounds exemplified by the success in describing doping-induced lattice parameter contraction. SCAN's description of the electronic structure of the parent compound is distinct from its preceding density-functional studies as SCAN predicts antiferromagnetic insulating ground state even in the multilayered compounds. The diminishing band gap is described by SCAN as a gradual and collective process contributed by Cu, Hg, and O components as opposed to the immediate process dominantly acted by the Hg states prescribed in previous studies [27]. We find this new physical description refreshing and hope it will encourage further advancement in the experimental techniques to synthesize the elusive parent compounds. Meanwhile, we note that doping these compounds with oxygen results in weaker magnetic moments, signalling an interplay between hole carrier concentrations in the $CuO_2$ planes with antiferromagnetic order. SCAN also correctly captures the magnetic inequivalence between copper planes for $n \geq 3$ observed in experiments. The DOS of doped compounds at $E_F$ can be significantly enhanced with some optimum excess oxygen concentration $x_o$, which manifests in the case of Hg-1201 as a peak in the normal-state, zero-temperature Sommerfeld parameter of the electronic specific heat. As the nature of this feature is currently of active interest, it is likely that modern first-principles density functional calculations can play an important role in unraveling the mysteries of quantum criticality.

## ACKNOWLEDGMENTS

The calculations were performed with the facilities of the Supercomputer Center, the Institute for Solid State Physics, the University of Tokyo, as well as the computational resource of Fujitsu PRIMERGY CX2550M5/CX2560M5(Oakbridge- CX) awarded by "Large-scale HPC Challenge" Project, Information Technology Center, The University of Tokyo.

## AUTHOR DECLARATIONS

**Conflict of Interest**

The authors have no conflicts to disclose.

Supplemental Material:

# First-principles electronic structure investigation of $HgBa_2Ca_{n-1}Cu_nO_{2n+2+x}$ with the SCAN density functional

Alpin N. Tatan[1,2], Jun Haruyama[2], and Osamu Sugino[1,2]

[1] Department of Physics, Graduate School of Science, The University of Tokyo, Tokyo 113-0033, Japan

[2] Institute for Solid State Physics, The University of Tokyo, Kashiwa, Chiba 277-8581, Japan

(Dated: May 4, 2022)

**Table S1**: Relaxed lattice parameters of $HgBa_2Ca_{n-1}Cu_nO_{2n+2+x}$ obtained by SCAN density functional calculation.

| $n$ | $a$ (Å) | | | | |
|---|---|---|---|---|---|
| | $x = 0$ | $x = 0.125$ | $x = 0.25$ | $x = 0.375$ | $x = 0.5$ |
| 1 | 3.897 | 3.858 | 3.842 | N/A | 3.8265 |
| 2 | 3.868 | 3.845 | 3.831 | 3.827 | 3.818 |
| 3 | 3.859 | 3.842 | 3.831 | 3.824 | 3.822 |
| 4 | 3.854 | N/A | | | |
| 5 | 3.851 | | | | |
| 6 | 3.849 | | | | |

| $n$ | $c$ (Å) | | | | |
|---|---|---|---|---|---|
| | $x = 0$ | $x = 0.125$ | $x = 0.25$ | $x = 0.375$ | $x = 0.5$ |
| 1 | 9.624 | 9.5954 | 9.5545 | N/A | 9.6204 |
| 2 | 12.839 | 12.793 | 12.739 | 12.762 | 12.782 |
| 3 | 16.039 | 15.971 | 15.895 | 15.910 | 15.978 |
| 4 | 19.188 | N/A | | | |
| 5 | 22.351 | | | | |
| 6 | 25.511 | | | | |

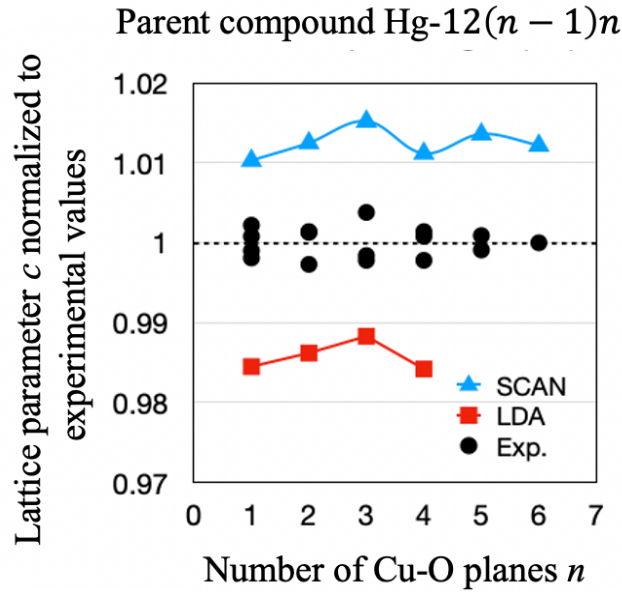

**Figure S1**: Lattice parameter $c$ of the parent compound Hg-12($n$ – 1)$n$ normalized to the average of experimental values for doped samples from multiple sources [S1 – S8]. SCAN's relaxed lattice parameters (blue triangles) for the parent compound are larger than the experimental values (dashed lines), in line with doping-induced lattice contraction. The smaller LDA results from Ref. [S9] (red squares) are included for comparison.

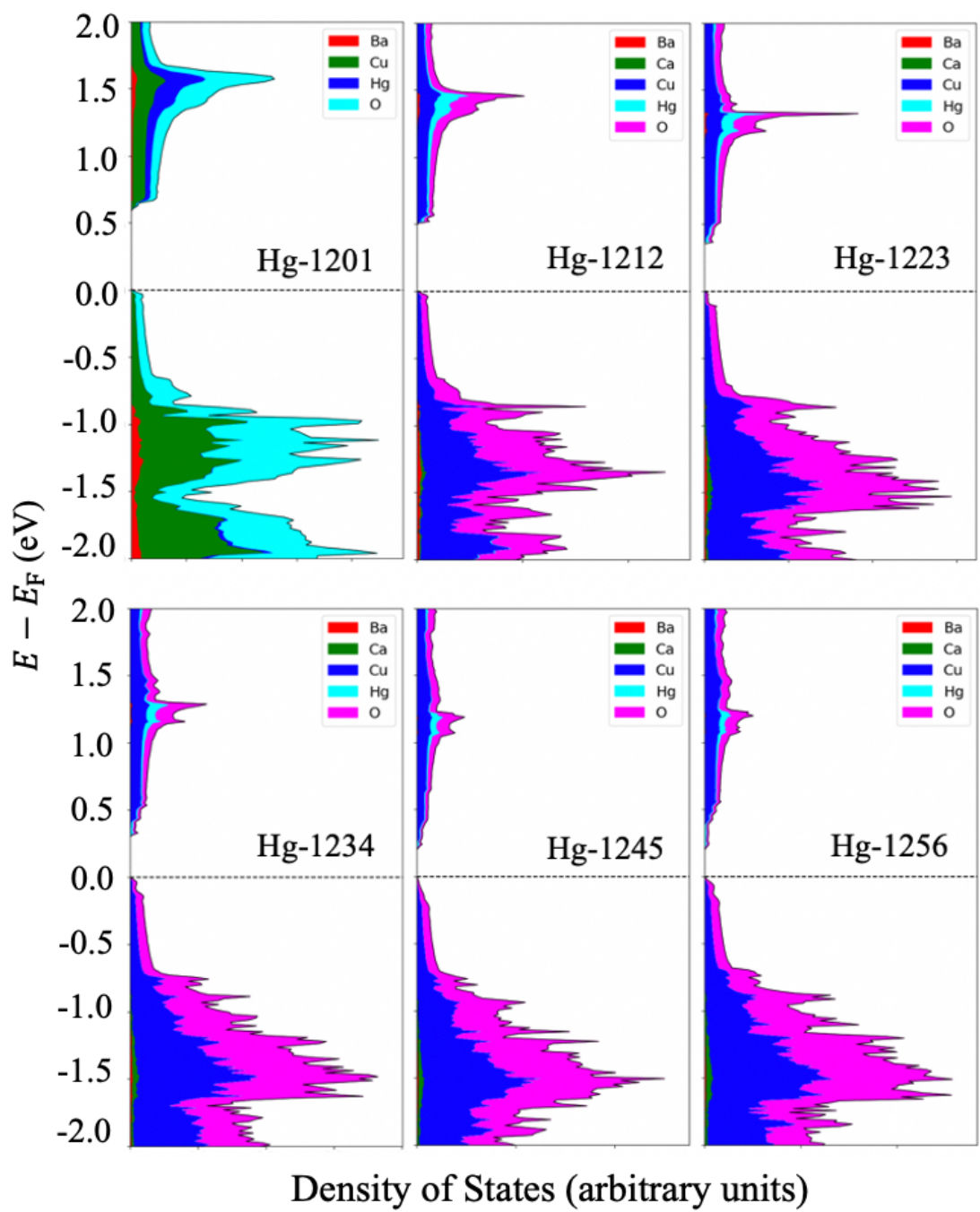

**Figure S2**: Stacked plots of the total density of states and its projections to each species of the parent compound Hg-12($n$ – 1)$n$.

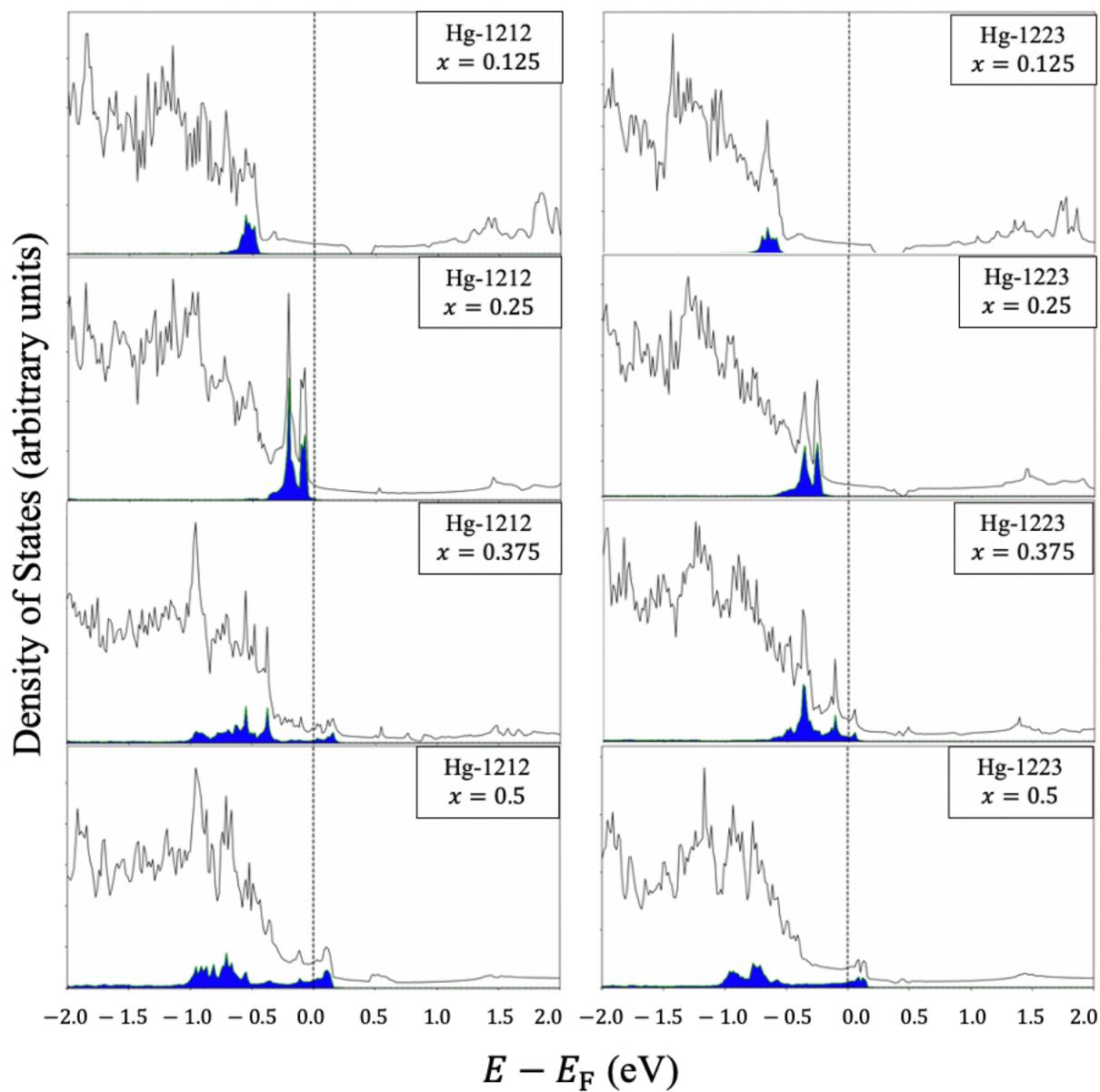

**Figure S3**: Total density of states (black traces) and the contributions from the dopant oxygen atom (blue area) for the bilayer Hg-1212 and trilayer Hg-1223 compounds.

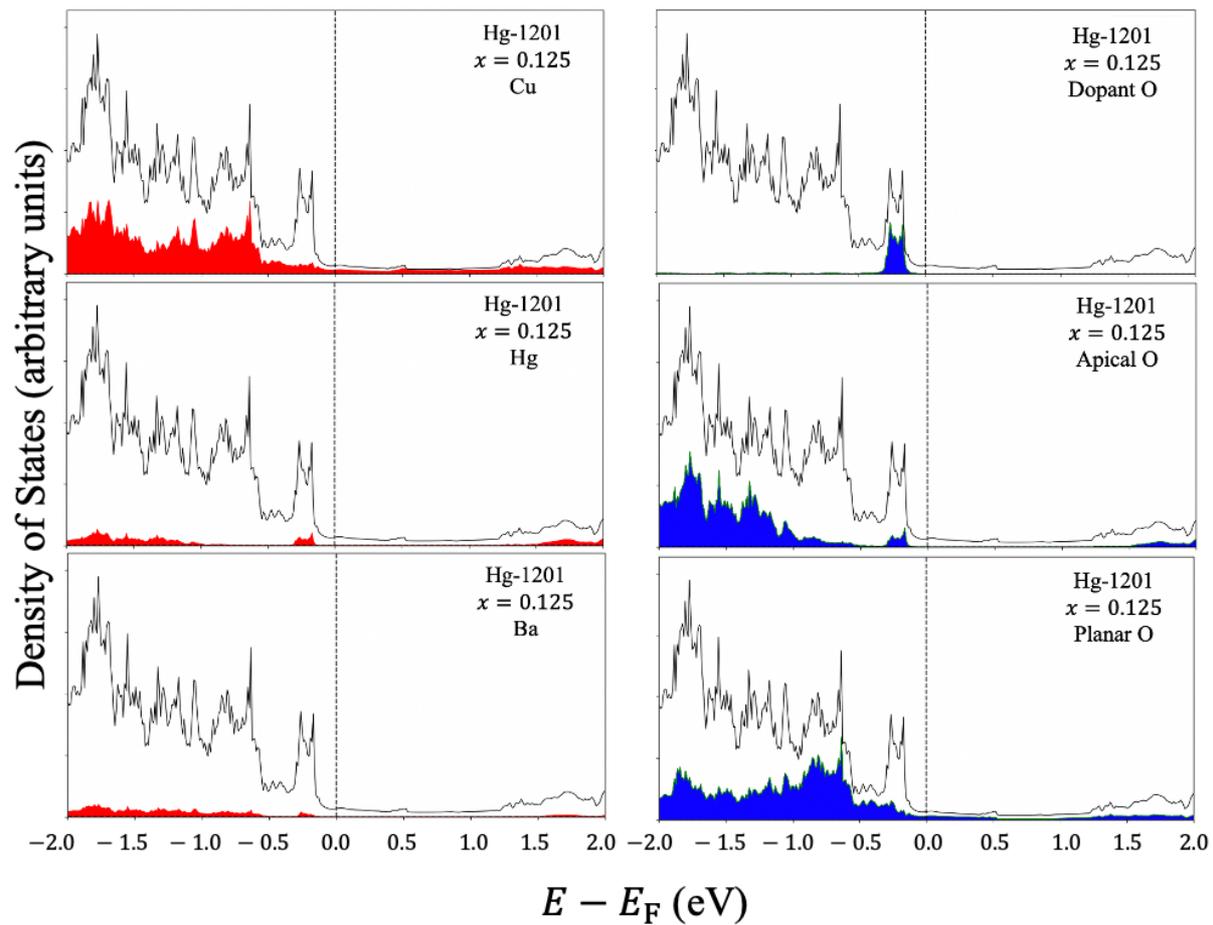

**Figure S4**: Total density of states (black traces) and the contributions from Cu, Hg, Ba (red area) and types of O atoms: dopant, apical, and planar (blue area) for Hg-1201 compound with $x = 0.125$.

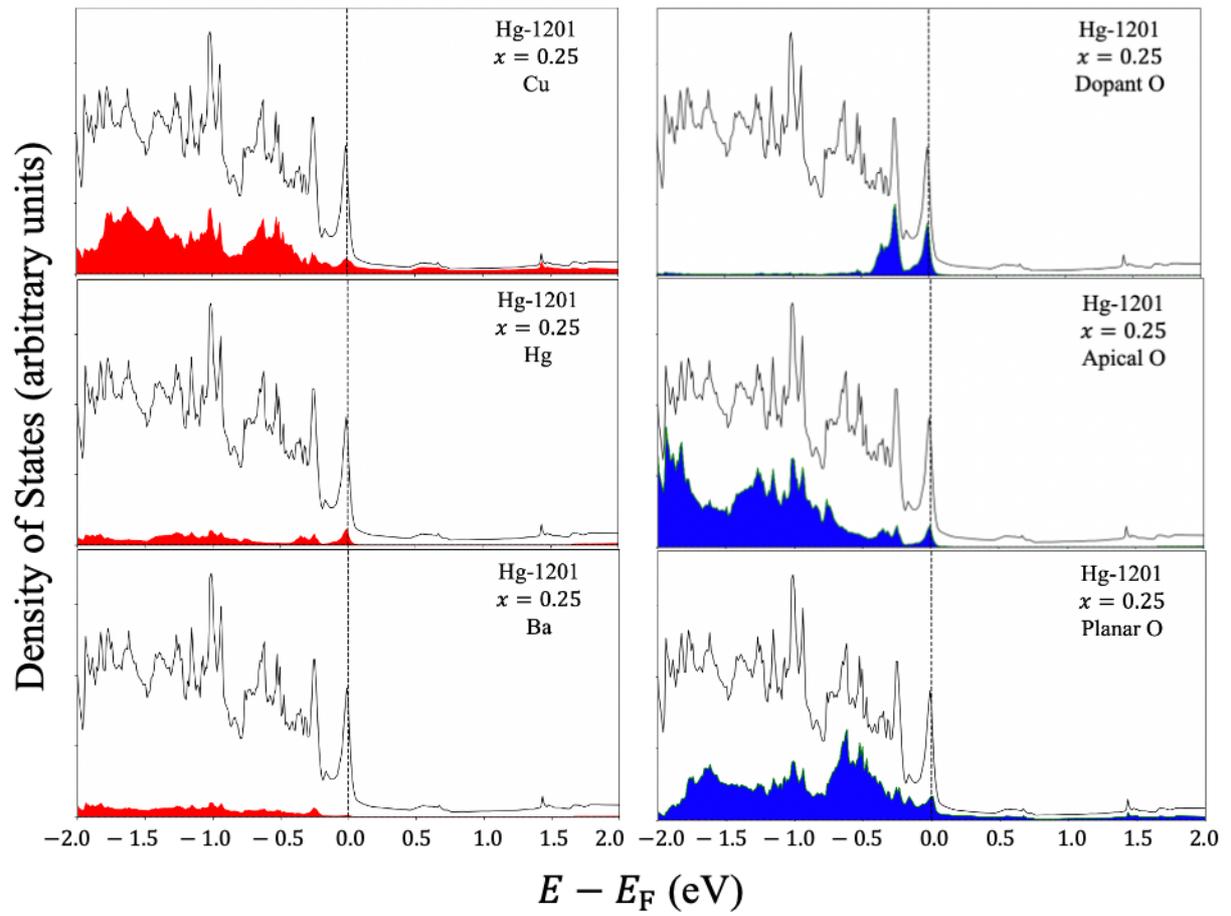

**Figure S5**: Total density of states (black traces) and the contributions from Cu, Hg, Ba (red area) and types of O atoms: dopant, apical, and planar (blue area) for Hg-1201 compound with $x = 0.25$.